\pdfoutput=1
\documentclass[lettersize,journal]{IEEEtran}
\usepackage{amsmath,amsfonts}
\usepackage{algorithmic}
\usepackage{algorithm}
\usepackage{array}
\usepackage[caption=false,font=normalsize,labelfont=sf,textfont=sf]{subfig}
\usepackage{textcomp}
\usepackage{stfloats}
\usepackage{url}
\usepackage{verbatim}
\usepackage{graphicx}
\usepackage{cite}
\usepackage{longtable}
\usepackage{hyperref}
\usepackage{color}
\usepackage{xcolor}
\usepackage{booktabs}
\newcommand{\tabitem}{~~\llap{\textbullet}~~}
\usepackage{soul}
\usepackage{braket}
\usepackage{marginnote}
\usepackage{authblk}
\begin{document}

\title{
  \Huge Towards a Unified Quantum Protocol Framework: \\
  \huge Classification, Implementation, and Use Cases}

\author[1,2]{Shraddha Singh *\thanks{shraddha.singh@yale.edu}}
\author[3]{Mina Doosti *\thanks{mdoosti@ed.ac.uk}}
\author[1,4]{Natansh Mathur}
\author[3,5]{Mahshid Delavar}
\author[1,6]{Atul Mantri}
\author[1,7]{Harold Ollivier}
\author[1,3]{Elham Kashefi}

\affil[1]{Laboratoire d'Informatique 6 (LIP6), Sorbonne Universit\'e, Paris, France}
\affil[2]{Department of Applied Physics, Yale University, New Haven, CT, USA}
\affil[3]{School of Informatics, University of Ediburgh, Edinburgh, UK}
\affil[4]{IRIF, Universit\'e Paris Cit\'e, CNRS \& QC Ware, Paris, France}
\affil[5]{Department of Computer Science, University of Warwick, Coventry, UK}
\affil[6]{QuISC, University of Maryland, Maryland, MD, USA}
\affil[7]{DI-ENS, PSL Research University, CNRS, INRIA, France}

\maketitle
\begin{abstract}
We present a framework for the unification and standardization of quantum network protocols, making their realization easier and expanding their use cases to a broader range of communities interested in quantum technologies. Our framework is available as an open-source repository, the Quantum Protocol Zoo. We follow a modular approach by identifying two key components: Functionality, which connects real-world applications; and Protocol, which is a set of instructions between two or many parties, at least one of which has a quantum device. Based on the different stages of the quantum internet and use-case in commercialization of quantum communication, our framework classifies quantum cryptographic functionalities and the various protocol designs implementing these functionalities. Towards this classification,  we introduce a novel concept of resource visualization for quantum protocols, which includes two interfaces: one to identify the building blocks for implementing a given protocol and another to identify accessible protocols when certain physical resources or functionalities are available. Such classification provides a hierarchy of quantum protocols based on their use-case and resource allocation. We have identified various valuable tools to improve its representation with a range of techniques, from abstract cryptography to graphical visualizations of the resource hierarchy in quantum networks. We elucidate the structure of the zoo and its primary features in this article to a broader class of quantum information scientists, physicists, computer science theorists and end-users. Since its introduction in 2018, the quantum protocol zoo has been a cornerstone in serving the quantum networks community in its ability to establish the use cases of emerging quantum internet networks. In that spirit, we also provide some of the applications of our framework from different perspectives. 
\end{abstract}


\section{Introduction}
Quantum Information Science (QIS) has progressed greatly since the first ideas of using quantum systems to process information about 40 years ago. QIS has found numerous applications in both computing and communication beyond the capabilities of classical systems~\cite{Fey82a, Wie83a, Fey84a,BB84a, Eke91a, Ben92a,Sho94a,Gro97b,JLP12a, JLP14a, JLP14b, JKL+18a,CRO+19a,VW20a}. This has provided hope to mankind for remarkable progress in computing, quantum networking, and quantum simulations, paving the way towards integrating quantum devices into commercial applications~\cite{dalzell2023quantum, lo2014secure, alleaume2014using}. Such a transition from purely academic development of quantum protocols and algorithms to industry-driven applications creates a gap between the existing body of knowledge accessible to different communities. A similar gap exists in the translation between theoretical works (in both quantum computation and quantum communication) and their experimental implementation. The need for a dedicated and well-structured knowledge repository in this field is paramount. Such repositories must also incorporate a formal representation and structure to bridge the aforementioned gaps with a scientific methodology while ensuring that algorithms and protocols are accessible to potential users, even if they are unfamiliar with the intricacies of physical and software implementations. Additionally, modularity in various quantum functionalities is essential, benefiting not only quantum software developers but also streamlining the development of commercially viable systems and applications, which can be captured only in a well-structured and modular framework. The formalization of knowledge and the creation of a formal resource theory for quantum networks and protocols are also critical from a theoretical perspective, as they provide a methodology for developing new applications in a more efficient and formal manner. Drawing inspiration from successful initiatives like the Complexity Zoo~\cite{compzoo}, the Error Correction Zoo~\cite{eczoo}, and the Quantum Algorithm Zoo~\cite{qalgzoo}, this endeavour seeks to comprehensively address these challenges in the field of quantum communication. While some surveys of existing quantum protocols exist~\cite{broadbent2016quantum}, \cite{pirandola2020advances}, \cite{kimble2008quantum}, \cite{sidhu2021advances}, \cite{singh2021quantum}, \cite{kumar2021state}, \cite{illiano2022quantum}, \cite{vidick2023introduction}, the absence of a formal categorization of knowledge in this field underscores the urgency of developing these frameworks and repositories.

In this paper, we present a unified framework for quantum network protocols. We provide a repository of quantum protocols, \emph{The Quantum Protocol Zoo}~\cite{zoo-all}, which is a compilation of various quantum network applications and protocols in the format of a wiki. It also provides a proper framework to categorize and analyze them in a structured manner. With the advent of quantum internet in the experimental stage, it becomes necessary to classify the existing set of protocols into different stages \cite{wehner2018quantum} based on their readiness for technological advancement from a primitive early stage (where we only have the capability to prepare and measure quantum states) to fully developed fault-tolerant quantum computing. The quantum protocol zoo offers this classification of protocols into various stacks of quantum internet - the \emph{hardware stack}, the \emph{software stack}, the \emph{application stack} and the \emph{user stack} -  to fill in the aforementioned gap. We also provide modular building blocks - in code and concept - to provide an implementation which is easy to integrate for the developers. Moreover, we identify a need for a unified theory to analyse the requirements for implementing a functionality or a protocol which could range from requiring high-level network protocols to real physical resources in order to provide a hierarchical structure to resources. 

\begin{figure}
    \centering
    \includegraphics[width=0.45\textwidth]{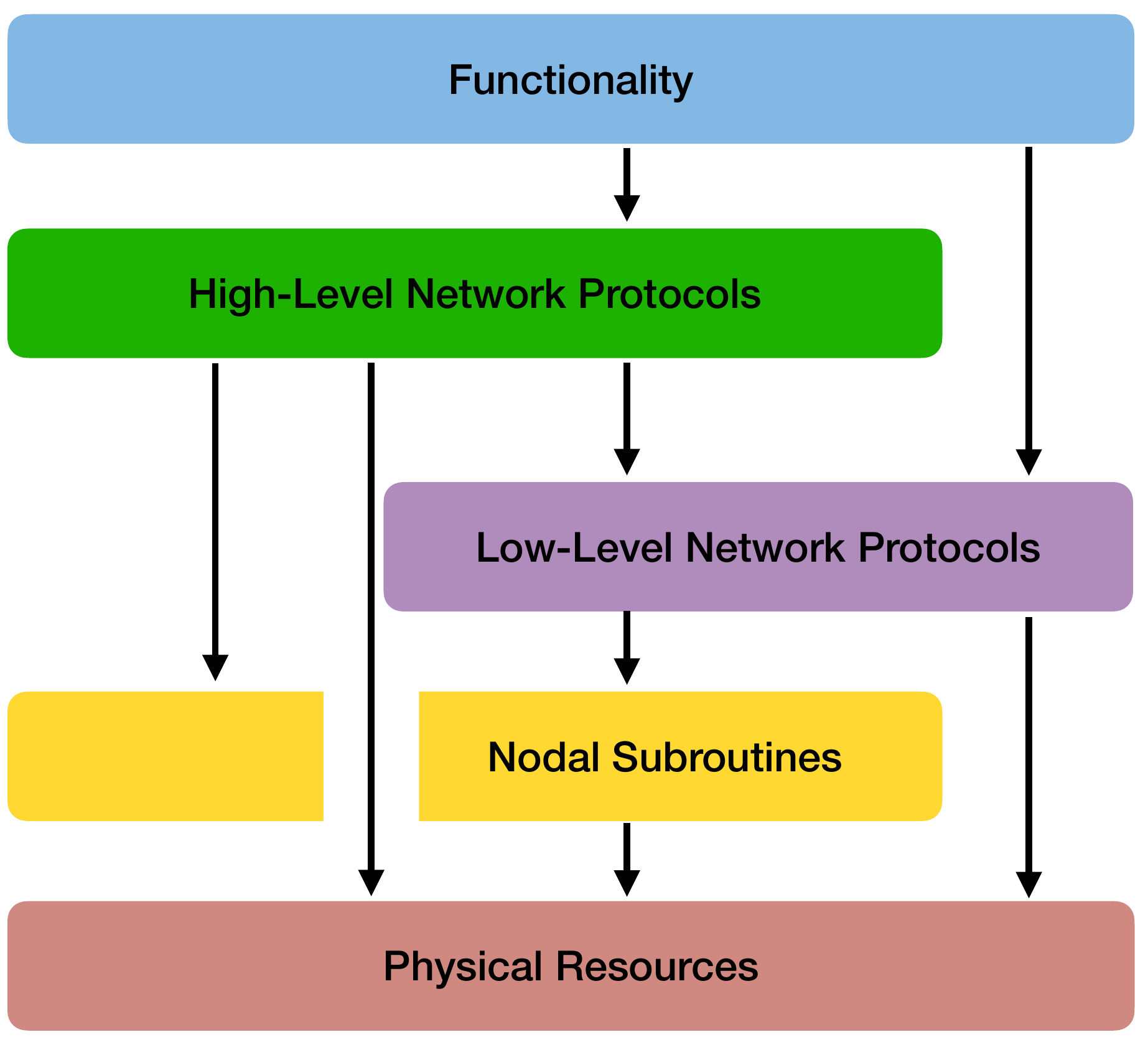}
    \caption{Resource hierarchy for quantum network protocols. The network stack represents a hierarchical abstraction of available quantum technology to quantum functionality. The arrows indicate the allowed paths between different stacks. Starting from the top, we descend from the desired \emph{Functionality} (representing a specific application) through various levels of protocols to the necessary \emph{Physical Resources} to realize it. The \emph{Functionality} stack represents various network applications such as digital signatures, cloud computing, and electronic voting, in other words, what is expected to be achieved through a quantum network.  In our proposed quantum network framework, these functionalities are implemented through quantum protocols, which can exist at any level within the stack, ranging from \emph{High-level Network Protocols} (protocols not used as sub-protocols), such as delegated quantum computing to \emph{Low-level Network Protocol} (which could be run as part of higher-level protocols), such as quantum key distribution.
    Furthermore, quantum protocols, regardless of their position in the hierarchy, may require parties to perform specific, often local, quantum or classical procedures, which we refer to as \emph{Nodal Subroutine}. Examples of nodal subroutines include quantum error correction. At the lowest level of the stack, we find the quantum technologies realized through the \emph{Physical Resources} required to implement these protocols and functionalities. They include single-qubit preparation and measurement, communication channels, and quantum gates. A bottom-up optimization in this stack is valid for small experimental labs to decide which functionalities are accessible to them given available quantum technologies. For a larger quantum network set-up, a top-down approach gives an estimate of the required quantum technologies for required quantum network applications.
}
    \label{fig:network-stack}
\end{figure}

We develop a resource hierarchy for quantum network protocols as depicted by Fig.~\ref{fig:network-stack}, with the aim of indicating the resourcefulness of the constituents of the different stacks shown. At the top of the stack lies functionality which connects quantum networks with real-world applications. The functionalities take place between two or more parties located at geographically different locations or nodes and communicate securely through quantum \emph{network protocols}. Thus, next in the hierarchy lies the stack of network protocols which can be used to achieve a specific functionality. Based on the resources required and the use case of the various protocols, this stack can be further subdivided. Here, for the protocols described in the quantum protocol zoo, we have identified two levels denoted by, \emph{Low-Level Protocols} and \emph{High-Level Protocols} where the former is sometimes embedded in the latter. The next stack comprizes some high-end \emph{nodal subroutines} which are quantum tasks specific to each node, for example, a random number generator or quantum error correction. Finally, the network links between nodes and the abilities to prepare-and-measure quantum states and apply quantum gate operations at each node are all combined and represented by the lowest node of \emph{Physical Resources}. In going from bottom to top in this hierarchy, we choose the functionalities based on the available quantum technology. The top-down approach on the other hand focuses on developing the required quantum technologies for specific functionalities in mind.

The quantum protocol zoo identifies and makes use of this hierarchy in various modes of representations to aid the connection between industry, theorists, software engineers and experimentalists. The rest of the paper explains the various aspects of the zoo as follows. Sec.~\ref{sec:framework} introduces our framework for the classification of quantum communication at both ends; functionalities as well as protocols. We then emphasize the need for modularity and present each functionality and the required classes of protocols with a textual (Sec.~\ref{sec:TextRep}) as well as graphical (Sec.~\ref{sec:KGRep}) representation, aimed at establishing a unified comprehension and implementation of quantum protocols. Sec.~\ref{sec:applications} underscores the implications and significance of our approach in representing quantum protocols, contributing to a well-organized knowledge repository for the quantum community. For example, the textual representation of the functionality highlights the `real world application and the current efficiency of the required quantum technologies' for the industries, while the protocols are given an `algorithmic representation' for the software library (see Sec.~\ref{sec:application-code}). We also quote the benchmark values for state-of-the-art experiments for different protocols for the experimentalists (see Sec.~\ref{sec:bench}). The protocol zoo can also serve as a live review of different quantum cryptographic functionalities, written for people with a diverse background and in different levels of accessibility. Finally, the graphical representation is extracted from the textual representation in the form of the \emph{Knowledge Graph} (KG) which presents the hierarchy in protocols abstracted in Fig.~\ref{fig:network-stack}. The KG presents a useful resource for theorists where we identify the embedding of various low-level protocols as subroutines to other high-level protocols. This representation identifies the building blocks of quantum networks which aid in the development of the abstract cryptography framework (see Sec.~\ref{sec:application-crypto}). The combination of different modes of knowledge representation (such as KG) and the modular structure of the zoo, guide towards a structured and secure integration of quantum functionalities to existing classical networks, which is a challenging task. It also helps in selecting the right quantum network architecture for specific functionality and vice versa (see Sec.~\ref{sec:application-network}). Additionally, we identify a new example of new protocol designs inspired by our framework in Sec.~\ref{sec:new-proto}. Finally, we conclude the paper in Sec.~\ref{sec:conclusion}.

\section{A framework for quantum protocols}\label{sec:framework}
We begin by introducing two crucial terminologies in our framework: Functionality and Protocol. Functionality is a user-end term which denotes real-world applications such as key distribution, client-server delegated quantum computing, digital signatures, and the list goes on. Protocol on the other hand is a formal set of steps to be taken in order to achieve the functionality between various nodes in the network. In the quantum case, some of these steps use quantum communication/computation to implement the desired functionality, which could be an application designed for a world before or after we have accomplished building a fully fault-tolerant quantum computer. This segregation was inspired by a modular approach towards the framework for representing network protocols. 

Before we dive into the framework, we motivate this need for modularity and thus, our chosen framework. Consider the application end-users that will invest in new infrastructure such as quantum internet not only because it offers them some advantage when the decision is taken but also if it continues to do so in the future. This usually means that an initial solution might be tested and later expanded as the need grows in terms of connection endpoints, transferred data, etc. This, in turn, imposes that new devices can be added to existing infrastructure to amortize the initial investment cost made at earlier deployment dates.
Similarly, developers in charge of integrating new services in existing communication stacks need to think in terms of functionalities. Depending on the exact context in which the functionality is used, one out of many possible protocols will be used, each using, in turn, some already implemented functionality. As a matter of fact, developers rely on modularity to extract commonalities in their code so that they group reusable portions in libraries, saving considerable effort and improving overall security and robustness. This modular methodology similarly applies to any network engineering task, where the network's configuration is typically established irrespective of its ultimate application. Assessing the components through benchmarking can be seen as a strategy to condense their performance into a handful of metrics, which offer valuable insights for anticipating their behaviour in real-world scenarios. Experimentalists also rely extensively on modularity as they explore and devise methods for manipulating quantum systems. The capacity to systematically assess the potential to achieve certain performance levels derived from their experimental findings - when transitioning their studied setups into network-integrated devices is of paramount importance. Without this modular evaluation, the motivation to conduct experiments would be notably diminished. Finally, for theoreticians, the development of a formal framework for discussing quantum protocols in terms of resources and with a modular perspective is highly important, motivated by several key factors in quantum information theory and cryptography. From a resource theory perspective, this framework enables a structured analysis of the distribution and transformation of quantum resources, providing insights into resource-efficient quantum protocols and applications. It also quantifies the ``quantumness" required for various tasks. Simultaneously, the modular perspective facilitates the construction of quantum systems and protocols from reusable, independently validated components, promoting faster development and greater accessibility. In quantum cryptography, theoreticians, particularly cryptographers, face a complex challenge. They must achieve modular behaviour without imposing restrictive assumptions on adversaries, which could undermine the security advantage of quantum cryptography. Fortunately, cryptographic frameworks like those presented in Refs.~\cite{Can01a, maurer2011abstract} introduce composability, allowing protocols to be used repeatedly and integrated into larger applications while maintaining security, even in the absence of assumptions about adversaries' behaviour. These frameworks address the need for robust security analysis, critical for preserving the unique advantage of quantum cryptography.

Recognizing modularity as a shared focal point among all categories of Protocol Zoo users, we decided to leverage this shared attribute as a fundamental strength in shaping the framework's design. The three crucial features of our framework are (1) Categorization (2) Textual Representation (3) Graphical Representation, and below we will discuss each of their roles in the zoo explicitly. In order to study the plethora of functionalities and the wide range of protocols implementing them, we came up with a classification to systematically study each component. This classification is covered in Sec.~\ref{sec:modular} where we also describe the motivation towards this modular approach of categorizing network protocols. The categorization of Functionality and Protocols inspired a textual representation, described in Sec.~\ref{sec:TextRep}, which covers a non-math summary for the non-quantum folks, a broad overview of efficiency for the industry, state-of-the-art benchmarked values and resource assumptions for experimentalists as well as pseudo code descriptions for the software library generators. The graphical representation, described in Sec.~\ref{sec:KGRep}, was then extracted from the myriads of textual description to analyze the resource hierarchy elaborated in Fig.~\ref{fig:network-stack}. The graphical representation is two-fold: the lineage graph and the resource graph. The knowledge graph aids the abstract cryptography framework by identifying the basic building blocks for which universal composability is a necessity. The resource graph on the other hand directs towards a resource theory of quantum networks.

\subsection{Categorization}\label{sec:modular}
\begin{figure}[!ht]
  \centering
  \includegraphics[width=
\linewidth]{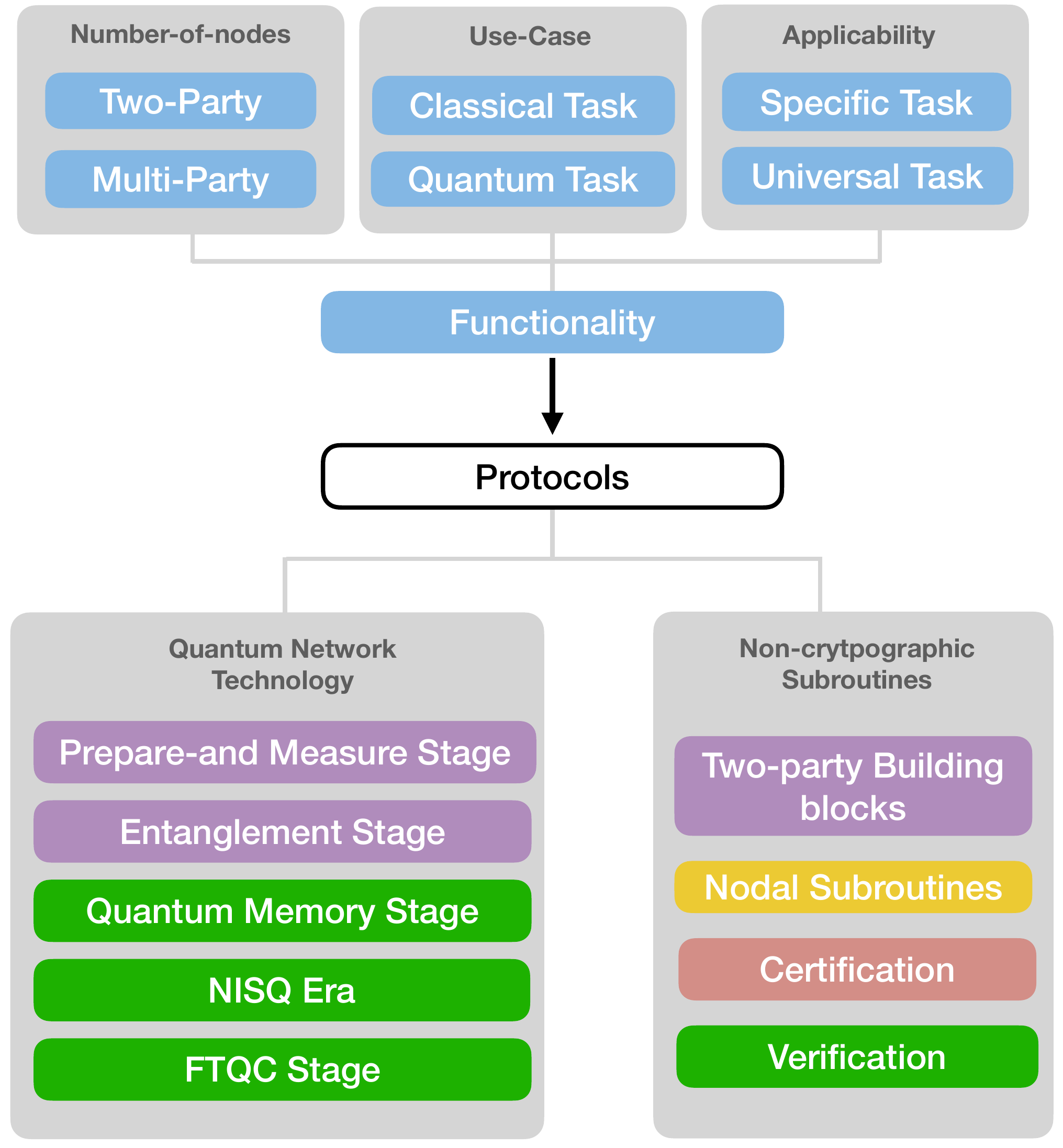}
  \caption{Different classes of categorization captured in the quantum protocol zoo using the colour coding of Fig.~\ref{fig:network-stack}. This figure represents the categorization of functionalities (as a high-level target application shown in Fig.~\ref{fig:network-stack}) based on different figures of merit, such as the required number of nodes for communication, quantum or classical use-case and generality of application. This categorization highlights the importance of the notion of \emph{quantum functionality} for a quantum network. For instance, the number of nodes determines the complexity of the quantum network. The use case determines whether the functionality achieves a classical task enhanced by quantum technologies or a quantum task with no analogue in the classical setting. Such categorization encourages comparison between the state-of-the-art functionality implementation in quantum and classical (or classical equipped with post-quantum cryptography) networks. Finally, the applicability decides whether the concerned functionality addresses a specific problem, like e-voting, or has a wide range of applications such as key-distribution and quantum authentication. Each quantum functionality can be implemented using different quantum protocols. The quantum protocols can also be distinguished on the basis of required quantum network resources and stage, as well as the complexity and availability of the nodal subroutines and physical resources that they use. Here, the acronyms NISQ and FTQC denote near-term intermediate quantum device and fault-tolerant quantum computing stages, respectively.
  }
  \label{fig:methodology}
\end{figure}

Categorizing functionality helps in ranking them on the basis of usefulness as indicated by their real-world applications at the user end. For example, some protocols are communication between two parties while some involve more than two parties. Not only does this apply stronger restrictions on security but also their physical implementation as more advanced quantum technologies become available. Another categorization specifies whether a functionality is truly quantum, that is, it does not exist in the world of classical computers or if it is a classical task which can be enhanced in terms of security or otherwise by quantum communication. For example, Blind Quantum Computing (BQC)~\cite{broadbent2015delegating} is a quantum functionality which hides the input, the output, as well as quantum computation or circuit of a minimally-abled client from a server with a fully fault-tolerant quantum computer. This functionality is not achievable so far in classical computers where the analogous classical scheme (Homomorphic Encryption) can only hide the input and output but not the computation from the server. Finally, an essential categorization is whether a functionality achieves a specific task of, say, Digital signatures, or a universal task, for example, delegated quantum computation between client and server. These categorizations can be used by industries as well as experimentalists to determine which type of applications they would like to focus on given the available quantum technology and requirements of society and research. This categorization favours the top-down approach of using functionalities to decide the course of research and commercialization in quantum networks.

We also categorize protocols which focus on the bottom-up approach of targeting specific functionalities based on the resources available. Here, we categorize protocols for each functionality based on the network stage they belong to, such that, it provides a clear path towards a quantum internet. In addition, we also classify the non-cryptographic protocols in the categories of nodal subroutines, for example: quantum error correction, two-party building blocks, like quantum teleportation, and certification and verification of quantum computers. We dedicate a textual description to each functionality and protocol, elaborated in the next section. 
\subsection{Textual representation} \label{sec:TextRep}
Inspired by the modular design discussed above, our framework includes a \emph{Functionality description} for each functionality, and the associated classes of protocols are also provided with a \emph{Protocol description}.
\subsubsection{Functionality structure} 
The Functionality structure, which can be found in Table~\ref{tab:functionality}, gives an overall picture of a class of quantum protocols achieving a common communication task. The main goal of the functionality page is to present a unified class of protocols achieving the same task in a quantum network, and their use-case in real-life applications. In addition, we specify the various available protocols classified on the basis of quantum technologies required by their implementation. We provide a comparison of the state-of-the-art implementation with the analogous classical or post-quantum secure technology in terms of security, speed and efficiency of communication. Such comparison highlights the components of the quantum implementation which will promote its commercialization: applicability, use-case, hardware implementation or security bounds. We refer to the functionality page for the class of quantum digital signature schemes~\cite{zooqds} as an example. 

\subsubsection{Protocol structure}
Each protocol in the protocol list of a functionality page opens up a new leaf in the repository, summarizing the associated protocol design. A protocol is a sequence of steps precisely specifying the actions that need to be run by two or more parties to accomplish the task described by the functionality. Our protocol structure can be found in Table~\ref{tab:protocol}. The goal of these pages is to provide a compact overview of different implementations of a functionality for available quantum technology. It highlights the main task achieved and the assumptions used to highlight the pre-requisites, outlines a no-math story describing the protocol for a quick review, a pseudo-code for the software stack, and finally all related theoretical modification and experimental implementations bench-marking the performance and security of the protocol. The protocol pages highlight the improvements required in specific implementations in terms of security bounds and experimental performance. We direct the users to the following example: a functionality page~\cite{zooqds} and a protocol page~\cite{zooqdsprotocol} where we use the above-mentioned structure for the Quantum Digital Signatures Scheme. 

\subsection{Knowledge Graph representation}\label{sec:KGRep}

In order to have a high-level view of each protocol and its requirements, as well as its relation to other protocols, we extract from the textual representation a \emph{knowledge graph}. The knowledge graph is represented graphically thanks to the open-source library visjs network~\cite{visjs}. 
\subsubsection{Protocol Decomposition Graph}
\begin{figure*}
    \centering
    \includegraphics[width=\textwidth]{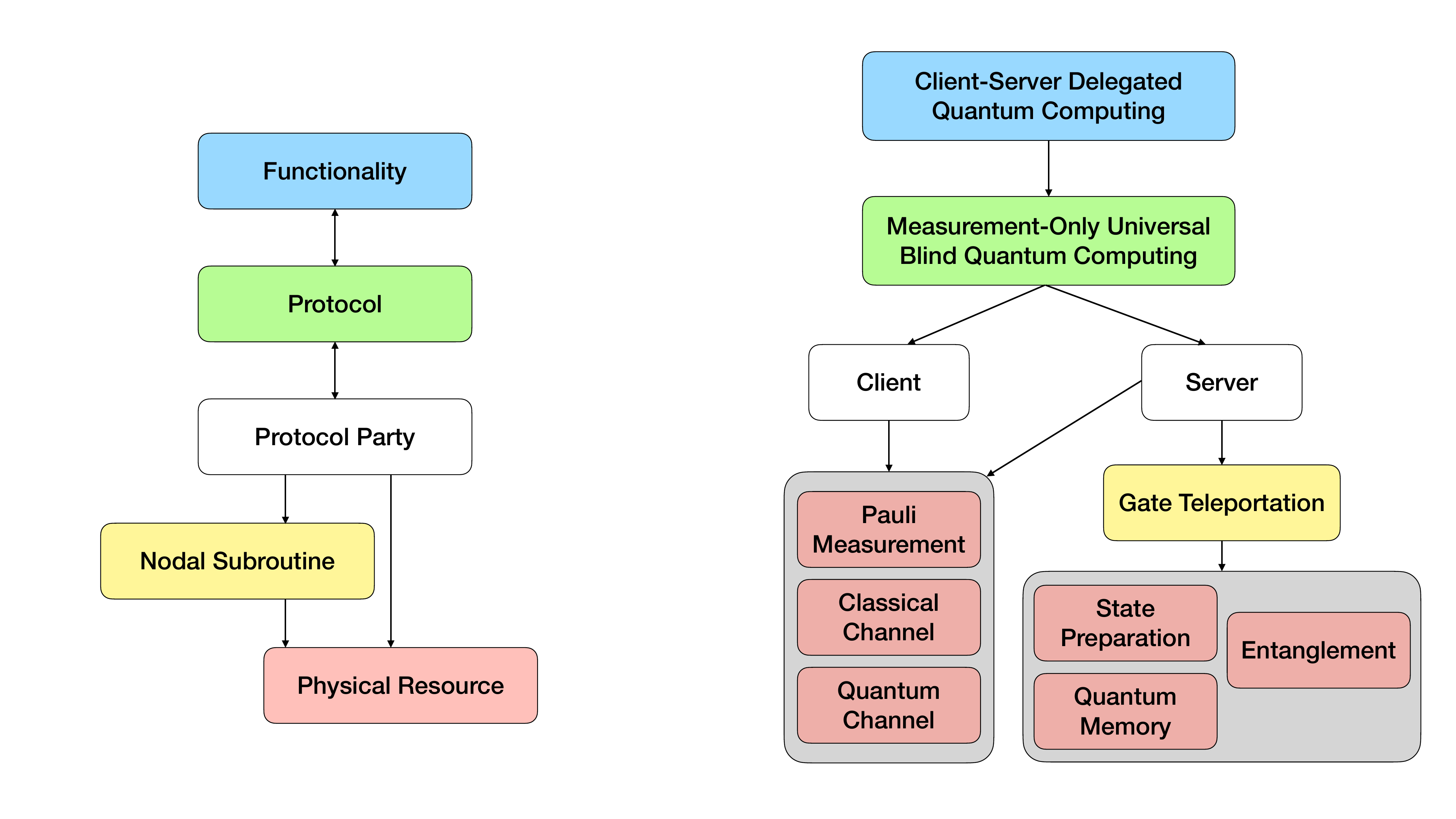}
    \caption{(Left) Protocol Decomposition Graph. Each protocol description consists of the above breakdown in terms of the parties involved and the network subroutines or resources required by each party. (Right) Example of asymmetric nodes (protocol parties) in decomposition graph. Measurement-Only Universal Blind Quantum Computing (UBQC) presents a protocol where a client with only single-qubit measurement abilities delegates its computation to a server with quantum memory and capabilities to securely do quantum gates, state preparation, and teleportation. Here, the client and server asymmetric are protocol parties in that the node representing the client can be a weak node where a fault-tolerant quantum computer does not need to be present. In the post-NISQ era, the server would also require quantum error correction and all other physical resources listed here.}
    \label{fig:kg}
\end{figure*}
A protocol decomposition in terms of the hierarchical network stack picture presented in Fig.~\ref{fig:network-stack} is extracted from the textual representation of each protocol. These decompositions are presented in terms of graphs with geographical nodes denoting the various protocol parties. A list of required physical resources and nodal subroutines is provided for each protocol party. This representation is helpful in recognizing protocols and functionalities with asymmetric nodes, for example, Delegated Quantum Computing (see Fig.~\ref{fig:kg}). The main links between the mentioned various network stacks are described as follows.
\begin{itemize}
\item \textbf{Functionality \(\longleftrightarrow\) Protocol} indicates the protocol used to implement the functionality. The protocol can have an inner hierarchy as discussed in Fig.~\ref{fig:network-stack}. The double-ended arrow indicates that certain lower-level functionalities may also be required to implement a protocol.
\item \textbf{Protocol \(\longleftrightarrow\) Protocol Party} indicates the different parties involved in the protocol. In an asymmetric protocol, each party is at a different network stage, connected to different sets of required physical resources. The double-ended arrow indicates that a party may also require another protocol for its implementation.
\item \textbf{Protocol Party \(\longrightarrow\) Nodal Subroutine} indicates that a protocol requires the implementation of the intended local subroutine to be run unanimously on or between all nodes.
\item \textbf{Protocol Party \(\longrightarrow\) Physical Resource} indicates that a protocol requires access to the specified physical resource to be physically implemented. Physical resources could also have a hierarchical structure where some elementary resources might be required with additional features to compose yet another physical resource, for example, a secure/authenticated quantum channel requires a quantum channel to begin with. However, we refrain from adding this hierarchy within the physical resources for simplicity.
\item \textbf{Nodal Subroutine \(\longrightarrow\) Physical Resource} indicates that a nodal subroutine requires access to the specified physical resources on the hardware level.
\end{itemize}
\subsubsection{Knowledge Graph}
\begin{figure*}
    \centering
    \includegraphics[width=\textwidth]{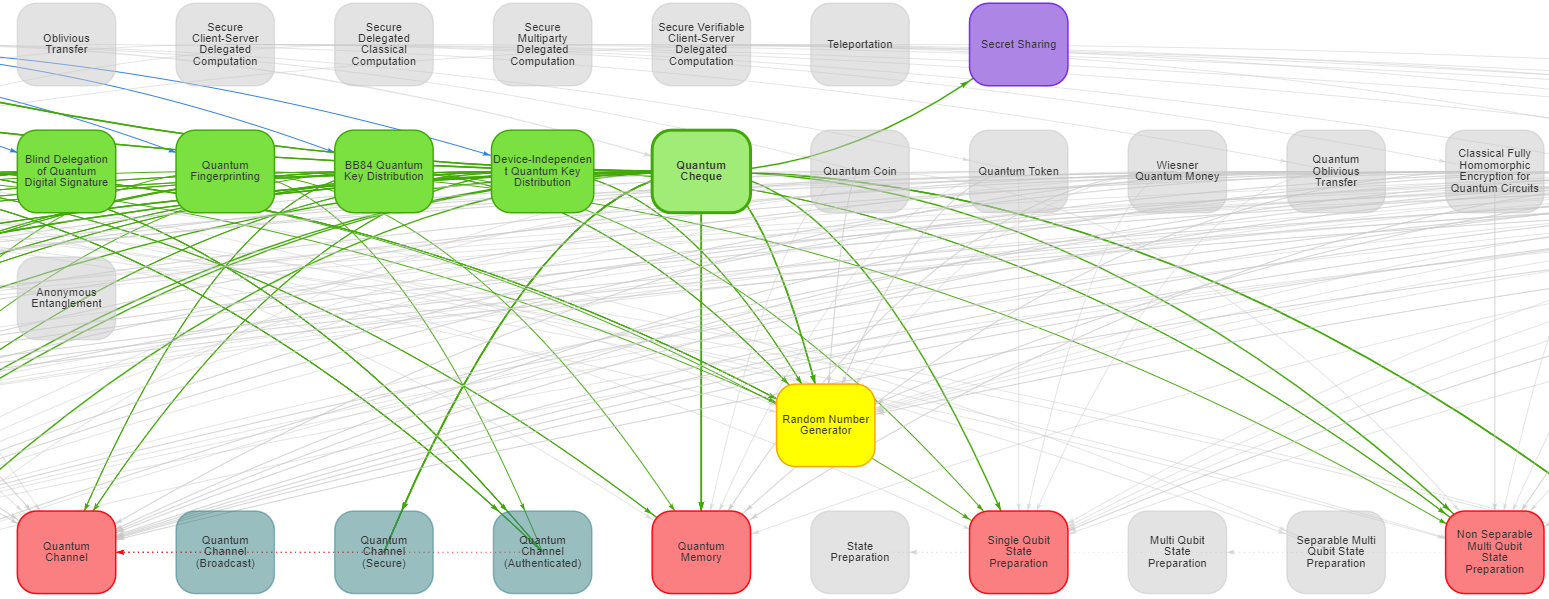}
    \caption{An example of the decomposition graph of a quantum protocol showing the relation to more than one quantum cryptographic functionality, demonstrated through the knowledge graph. In this example, \emph{Quantum Cheque} protocol~\cite{zooqcheque} is connected to different functionalities: digital signature~\cite{zooqds}, key distribution~\cite{zooqkd}, secret sharing functionality~\cite{zooss} and fingerprinting.}
    \label{fig:quantum_cheques}
\end{figure*}

This graph represents a unified picture of quantum network protocols using the \emph{protocol decomposition graph} obtained from the textual representation. The various network protocols are linked via common subroutines and low-level cryptographic protocols (see Fig.~\ref{fig:quantum_cheques}). \emph{The knowledge graph}, implemented for the protocols that have been added to the protocol zoo, is available at~\cite{zookg}. To better present the relationship between different nodes in the graph, two types of visualization called \emph{lineage} and \emph{resource} visualizations are proposed. The former highlights all ascendant and descendant nodes of the specific node that the user is interested in, while the latter highlights all the descendant nodes. The highlighting of the ``lineage" and ``resources" visualizations is obtained by traversing the graph with two dedicated in-browser algorithms. The lineage algorithm selects all ascendants and descendants of a given protocol. The resources algorithm starts with the selected physical resources and stores them in the list of available nodes. Since these nodes are available, their descendants must be available as well. Consequently, it augments the list of available nodes with all descendants of the selected nodes. Then, each direct ascendant of available nodes is examined. If all its descendants are in the available nodes list, it becomes available. If not, it remains unavailable. Doing so recursively allows construction of all the nodes that are available starting from the selected ones. The termination is guaranteed because the knowledge graph is acyclic. All algorithms associated with these visualizations have been implemented through in-browser code.

With these tools at hand, we want to facilitate protocol explorations and decision-making as to which protocols to use for proof of concept experiments or to define as a useful intermediate step towards building more complex and integrated applications.

\section{Applications and Impacts}\label{sec:applications}
An organized body of knowledge can have a fast and broad impact on the field, particularly when standards start to emerge, the community grows and people specialize, sometimes making it difficult to grasp a broad view of what is at stake. Now, we elicit the benefits of defining, enriching and enlarging the unified picture of the quantum network presented in the form of the quantum protocol zoo. In this section, we will specifically elaborate how this work champions the following aspects of quantum internet: (\ref{sec:bench}) Hardware efficiency in quantum networks, (\ref{sec:application-network}) Optimality in network architecture from the most primitive devices, (\ref{sec:application-code}) Software protocol library that enhances future scalability of quantum protocols, (\ref{sec:application-crypto}) Security analysis, especially in the context of composable security, (\ref{sec:new-proto}) New protocol designs where the deconstruction/reconstruction approach~\footnote{ It involves breaking down a complex problem or system, in our context, a quantum functionality into smaller, more manageable components, analyzing each component separately and then putting them back together to understand the overall behaviour or performance.}  can easily pinpoint bottlenecks in terms of functionality and performance.

\subsection{Roadmap for experimental and industrial advancements}\label{sec:bench}
The zoo can serve to establish a roadmap for benchmarking experimental advancements as well as industrial use cases. This can be accomplished by tracking state-of-the-art implementations for the various subroutines used in a protocol providing a given functionality. This then allows us to assess the global performance metrics for a functionality and to possibly compare it to alternative implementations such as their optimal classical counterparts. Such an approach was taken for instance in~\cite{LBSK22:benchmarking} and qualitatively highlights the readiness of quantum protocols for end-users.  But besides driving attention to protocols and functionalities that are either close to or very far from meeting end-user expectations, the modularity of the protocols can be used to assess quantitatively the effort that needs to be put to have satisfactory implementations. Such methodology was developed for instance in~\cite{A23:shortcuts,SWA23:requirements} by considering the minimal performance requirements of each physical resource used in a quantum protocol, provided all others were at their optimal values. While this only gives a lower bound on the required performance improvement, it provides a quantitative measure. Computing or simulating the protocols using their modular structure can thereby give a very clear view to experimentalists of which actions can be quick wins and where roadblocks can be.

As experimental capabilities progress, global initiatives focus on developing versatile devices for quantum networks, necessitating the design of specific operating systems to ensure device interoperability. Overcoming challenges related to the limitations of these networks compared to classical counterparts is crucial. Embedding higher-level functionalities into physical devices may help minimize latency and back-and-forth communication between series of commands, and also mitigate the impact of decoherence. The above-discussed concept of modularization serves as an initial step towards hardware-independent task allocation, aiding both software stack engineers and developers in writing enduring code for quantum protocol implementation.
This modularization of tasks is executed in a manner that strives for hardware independence to the greatest extent possible.
This is because protocols are developed and proven secure with few references to the actual platforms that they might be run on.
As a consequence, their description is stable in time.
On the contrary, the software stacks offering APIs to hardware or simulation platforms still tend to undergo rapid changes as a consequence of technological advances.
Our decomposition of algorithms into simpler resources could be a guide for both engineers developing the software stack and for code developers wanting to write long-lasting code for implementing quantum protocols.

In this direction, an outcome of the modular design of the zoo inspired the PyPi project~\cite{O21:qpz-atomics} interfacing two quantum network simulation platforms and providing a set of \emph{atomic functions} to be used to write platform agnostic code.
The process leading to such a library was to review protocols with the aim of collecting subroutines and resources needed to implement them.  
Each collected element was then ranked by its importance with regards to how many protocols or higher level functionality was depending on it.  
This led to specifications for the most important atomic functions selected according to this metric.
Along the way, we derived a slightly different graphical representation that focused on the type of nodes (atomic function vs.  
protocol).  
This allowed us to also assess the network stage that each of these functions belongs to and the interest in including the functionality in the library.
Further details about the extracted atomic functions and the derived graphical representations are presented in appendix~\ref{apx:atomic}.

We believe that this kind of approach, while still preliminary, should be generalized and will bear fruits as the number of protocols increases and as the need for simulation and benchmarking will inevitably require sustained coding and software integration efforts.

\subsection{Network Architecture}\label{sec:application-network}

As the quantum internet unfolds, it becomes increasingly apparent that the efficient construction and integration of networks~\cite{wehner2018quantum}, \cite{kimble2008quantum}, \cite{cacciapuoti2019quantum} are crucial steps. Achieving this objective demands a strategic optimization approach that carefully considers the intricacies of both hardware and protocols. It is essential to address the challenges arising from the limitations of current quantum hardware and the specific demands of quantum protocols~\cite{dahlberg2019link,satoh2020attacking}. In this context, drawing an analogy to the specialized classical network architectures designed for IoT sensors~\cite{sarwesh2017energy} proves insightful. By aligning the quantum network architecture with its intended applications, it becomes possible to enhance the overall performance. This approach holds particular promise for early-stage imperfect quantum devices grappling with stringent hardware limitations. 
It is also worth noting that quantum networks, despite their revolutionary potential, intricately rely on classical networks for many applications. Various classical network attributes like communication capacity, communication fidelity, and security of such classical networks play a crucial role in many quantum protocols. When selecting a quantum network architecture, it is paramount to keep this integration in mind during the analysis. offer invaluable assistance in addressing this intricate fusion of quantum and classical network elements, ensuring a seamless and effective coexistence. Our structured textual representation of quantum protocols provides the main requirements of these protocols, also, the main components of each quantum protocol are visualized through our graphical representation. The knowledge graph leads to finding the atomic tasks discussed in Sec.~\ref{sec:bench} which shows the common components among several protocols.
Thus, the design of quantum network architecture based on atomic tasks provides a quantum network supporting the maximum number of applications.

However, unlike classical networks, quantum network architectures can be also based on several fundamentally different ways of connecting nodes.  
While the quantum network stage that is capable of realizing only QKD schemes most resemble classical networks where photons directly carry information, others will be entanglement-based.  
The latter scheme overcomes the intrinsic risk of losing precious information by pre-sharing entanglement between the nodes.  
These networks could then achieve a higher level of universality than the QKD networks, as they can in principle run any protocol.  

Such universality is the mantra behind most current bottom-up approaches where the underlying architecture is developed first, and then, the possible applications compatible with this architecture are investigated \cite{dahlberg2019link}. 
Yet, universality is mostly a theoretical concept, as what motivates users is usually the performance of a smaller group of applications of interest to them.
In this perspective, designing and optimizing the bottlenecks of the architecture to deliver more of what users want is starting to be a major objective.
This can be achieved by taking a closer look at the required topology of the network or at ways to establish connections between nodes. For example, some form of secure multi-party computation requires a complete network while others are compatible with the simpler star topology. On the other hand, some protocols would naturally favour point-to-point link layers, while others would benefit greatly from sharing graph states instead of bipartite entangled states. There again, our proposed framework and its visualization tools greatly facilitate the task of grouping protocols having one or several technical characteristics in common.

\subsection{Establishing a Software Library}
\label{sec:application-code}

On the software front, considerable efforts have been dedicated to making quantum technology accessible through simulators and software packages, which aid in understanding imperfections, testing algorithms, and supporting education. While network-specific simulators are limited (Netsquid~\cite{CKDM+21:netsquid}, SimulaQron~\cite{DW18:simulaqron}, QuiSP~\cite{SHBN+21:quisp} and QuNetSim~\cite{D19:qunetsim}), most protocols require local quantum computing capabilities available through various quantum simulators like IBM Qiskit, Rigetti Forest, Microsoft Q\#, Google Cirq, Zapata Orquestra, and more. Furthermore, groups worldwide are developing quantum network proof-of-concept devices, driving research into operating systems that ensure interoperability and expose essential functionalities in the software stack.

Our proposed approach, put forward in the quantum protocol zoo, for decomposing the protocol into functionalities, subprotocols and ultimately into physical resources brings a useful framework for tackling part of these tasks. Unlike classical programming, developing quantum algorithms and protocols is currently in the early stages, where very few application-level libraries exist. The consequence is twofold: First, quantum researchers and engineers need to code from scratch many boilerplate functionalities, thus further increasing the entry costs for studying and developing more advanced protocols while being an error-prone process; second, because of its narrow user base, there is practically no user-based or experience-based guidance about the suitable structure of simulation tools. The decomposition gives tools to address these two problems in a natural way: The decomposition of protocols being based on cryptographic principles with composability in mind readily proposes a modularization of the protocol that could be readily replicated by code. Recognizing that some of these functionalities are used in several protocols yields utility function libraries. As mentioned earlier, debugging the codes for quantum protocols is a hard task. The intermediate state of quantum registers can be impractical to check by hand as well as automatically because of the large size of the classical representation of the state of these registers. Some quantum simulators do not even give access to the state. It is thus of paramount importance to develop unit tests that can be composed with one another in order to take simple functionality assertions and get the validity of the implementation at a higher level. Here again, the decomposition in terms of cryptographic functionalities provides ready-made tests consisting of asserting the correctness of the functionality. This approach is being used for implementing the above-mentioned library where unit tests are performed with the help of the \texttt{hypothesis} package for Python (which generates examples automatically, thus reducing the design of tests to really writing down the assertion for correctness).

And finally, interoperability can be achieved relatively simply because protocols are decomposed down to physical resources. As a matter of fact, simulation software as well as devices can always be thought of as providing (or simulating) physical resources. Mapping these physical resources to their device-independent version allows the construction of device and simulation software agnostic of the utility libraries.

\subsection{Analyzing the security of protocols}\label{sec:application-crypto}
\begin{figure*}[!ht]
    \centering
    \includegraphics[width=0.7\textwidth]{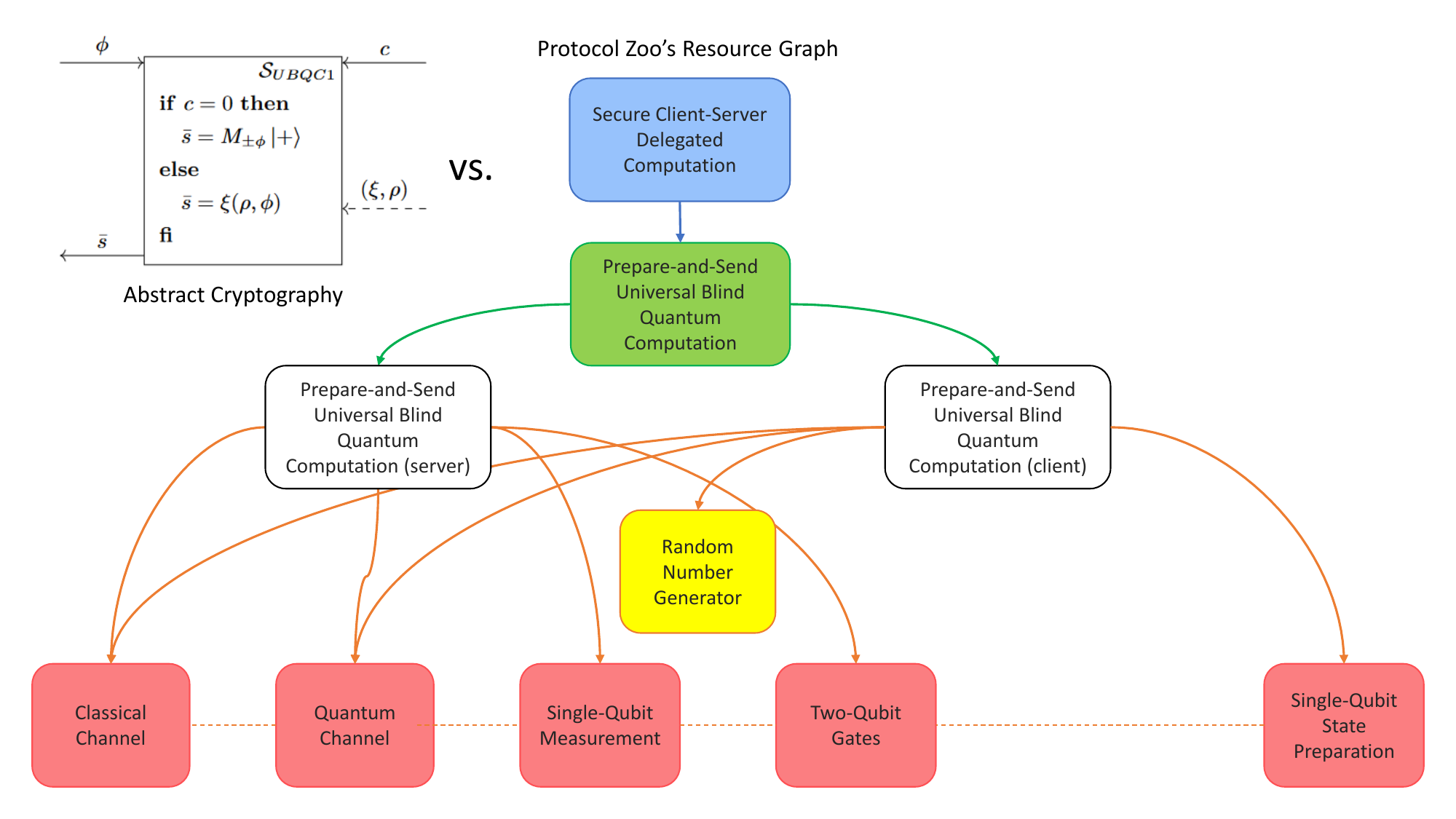}
    \caption{The comparison between the graphical representation of Prepare-and-Send Universal Blind Quantum Computation~\cite{zoopsubqs} adapted from the \emph{resource graph} of quantum protocol zoo~\cite{zookg} (right) and its ideal functionality from abstract cryptography (left). In the ideal functionality, the right-hand side is the server and the left-hand side is the client who sends a single-qubit state and receives the output of a computation from the server given by $\overline{s}$ by measuring in the proper basis. The parameter $c$ characterises if the server is honest ($c=0$) or malicious ($c=1$), in which case the server deviation is denoted via $(\mathcal{E}, \rho)$, where $\mathcal{E}$ is a quantum channel and $\rho$ is a general quantum state. Equivalently from the protocol zoo resource graph, it can be seen that the resources on the client side are the ability to prepare a single qubit and send it via a quantum channel, in addition to having access to a classical channel. On the server side, single-qubit measurement and double-qubit gate resources are needed.}
    \label{fig:pzoo-ubqc}
\end{figure*}
In the cryptography literature, the design of complex cryptographic protocols such as e-voting and proving their security is considered a challenging task. The protocol designers usually combine simpler protocols and schemes in a proper way to design complex protocols \cite{camenisch2019multi}. Also, to prove the security of such protocols, they use a series of definitional frameworks like universal composability (UC) framework \cite{canetti2001universally} or Abstract Cryptography (AC) \cite{maurer2011abstract} aiming to modularize the security proofs via the use of general composition theorems. In these frameworks, first, the security of all building block schemes and protocols is proved or their ideal functionality is provided individually, then, the security of the complex protocol built by these schemes is analysed. Note that, today, the modular approach to security analysis of complex protocols has been widely accepted as a desirable security goal since it allows for the use of a protocol as a building block in other protocols.

Inspired by the notion of compositional frameworks, we distinguished the functionality from the protocol which has been reflected in our proposed textual representation. Although we are not using the same syntax as used in cryptography literature to define the ideal functionalities due to realisability by different communities, our framework still gives an idea of which functionality is expected from a set of protocols.

Also, as introduced in Sec.\ref{sec:KGRep}, our graphical representation shows the relations between different functionalities and protocols. We believe the knowledge graph helps cryptographers easily find the building blocks of complex quantum protocols and use any compositional security frameworks to analyse their security. For example, the decomposition graph of ``quantum cheque" shows it is combined with four functionalities: digital signature, key distribution, secret sharing and fingerprinting (Fig.\ref{fig:quantum_cheques}). Thus, one may only need to provide the ideal case of these functionalities and use the AC or UC framework to prove the security of quantum cheque in a hybrid model \cite{canetti2001universally}.

We can show through the following example, how our graphical representation is linked to the ideal and real functionalities of a protocol in the AC framework. Prepare-and-Send Universal Blind Quantum Computation \cite{broadbent2009universal} is a  protocol that achieves the functionality of \emph{Secure Client-Server Delegated Quantum Computation}. The composable security of this protocol has been investigated in the AC framework~\cite{badertscher2020security}, where the ideal functionality has been defined by Fig.~\ref{fig:pzoo-ubqc} (left). In this ideal functionality, the right-hand side is the server and the left-hand side is the client who sends a single $\ket{+_{\phi}}$ qubit $\phi \in \{0,\frac{\pi}{4}, ..., \frac{7\pi}{4}\}$ and receives the output of a computation from the server given by $\overline{s}$ by measuring in $\{\ket{+_{\phi}}, \ket{-_{\phi}}\}$ basis. The parameter $c$ characterises if the server is honest ($c=0$) or malicious ($c=1$) in which case the server deviation is denoted via $(\mathcal{E}, \rho)$, where $\mathcal{E}$ is a quantum channel (CPTP map) and $\rho$ is a general quantum state. Now let us look at the graphical representation of the protocol demonstrated in Fig.~\ref{fig:pzoo-ubqc} (right). 
As can be seen, each party has been connected to its underlying resources. The resources on the client side are the ability to prepare a single qubit and send it via a quantum channel, in addition to having access to a classical channel. The server, on the other hand, needs to be able to perform a single qubit measurement and double qubit gates, to be able to run the universal computation. We would like to emphasise that excluding the malicious scenario, the ideal functionality can be intuitively related to the graphical representation without even going through the details of the protocol. In addition, the graphical representation recognises the real protocol. One can see that the parameter $\phi$, in the ideal functionality, will be a single qubit state in the real protocol, sent via a quantum channel from the client to the server on a shared quantum channel between them. This example demonstrates how the provided framework of protocol zoo can facilitate the abstraction of an ideal functionality from a specific protocol, which is an important step towards studying them in a composable framework such as AC that enables rigorous security analysis of complicated quantum protocols.

Moreover, cryptographers can use the knowledge graph to find the most common protocols that are used as a building block of other protocols and prove their security in compositional frameworks. Once they prove the security of the common sub-protocol in these frameworks, the protocol designers will be sure that they can use the protocol in composition with other protocols to build more complex protocols without compromising security, simplifying the process of designing secure complex protocols.

\subsection{Developing new protocols}
\label{sec:new-proto}
Research on quantum networks has several aspects, one of which is the development of new protocols. With maturity advancing, a lot of interesting cryptographic primitives have been described in the literature and some are already included in the quantum protocol zoo functionalities or will be in the future. Some have several implementations, others a single one. More often than not, however, protocols can be improved in various aspects. Their optimization to better cope with limited hardware, integrate new capabilities, or improve their performance is an important aspect of the development of future quantum technologies.

As a consequence, there is a constant need for a back-and-forth shift of focus between designing and proving the security of protocols.  To this end, our framework facilitates the task of researching and improving protocols. The reason is twofold: first, it does identify sub-protocols that can sometimes be replaced with different variants, or that can be modified to acquire a new capability. Second, the decomposition of the protocol being compatible with composable security frameworks such as abstract cryptography, makes the proof of security for the modified protocol readily available.

This approach has been exploited in e.g.~\cite{KKLM+22:framework} where the verifiable blind quantum computation (VBQC) protocol was abstracted in a way that made it possible to identify three properties -- detection, insensitivity and correctness -- required to obtain security in delegated computations. The abstract protocol could then be used to propose design rules for concrete protocols that could then be tailored to take into account e.g. noisy hardware, provide additional functionality or encompass different situations altogether such as multiparty computation~\cite{KKLMO23} or the delegation of fault-tolerant computations~\cite{KLMO23}. More precisely, the quantum protocol zoo identifies the need for an optimised implementation of the subroutine for state preparation before the computation part can really begin in the VBQC protocols. 
One such improvement is the ability to perform the state preparation in a multi-client setting while retaining the security of the state preparation as demonstrated in \cite{KKLMO23}. Furthermore, since the modular approach of the zoo is compatible with the composable cryptographic frameworks, the security of the new VBQC protocol was obtained as a consequence of the composable security of the modified state preparation.

Such schema is not unique. To continue with secure multiparty protocols, \cite{DGJ+19a} has improved on \cite{CGS02a} by modifying implementations of the gates using a garbling approach and then proving secure composability before assembling each gate into a whole circuit. Similarly, quantum secure multi-party computing can be obtained from verifiable secret sharing (VSS) and as such, any new quantum protocol for this will potentially offer a way to improve quantum secure multi-party computing.

As these examples exhibit, the process of developing new algorithms often consists of improving smaller parts of protocols. In this respect, the protocol zoo identifies these various parts and helps assess what is potentially affected if a given functionality is improved or made accessible, while potentially easing the proof of security in the case the replacement parts can be shown to be composable. 

\section{Conclusion}\label{sec:conclusion}
In this paper, we proposed a framework for describing quantum protocols which provides a new and more informative approach to represent a quantum communication scheme by separating the notion of functionality and protocol. Furthermore, each functionality and protocol is categorized on the basis of applicability and resource requirements, respectively. We provide both textual and graphical representations which are aimed at championing the standardization of quantum protocol designs.  Finally, we discuss the applicability and benefits of our proposed framework in different domains and for different target communities and are hopeful that it facilitates the study and development of quantum protocols which in turn contribute towards further utilities of the current and future quantum networks. Our framework identifies a resource hierarchy in the quantum network protocols which aids advancement in quantum network research at all fronts, composable security analysis, building a software library, commercialization of quantum network protocols and experimental advancements.

As a concrete future direction, we propose a further formalization of this structure using resource theory, which will assist in the progress of various research directions such as abstract cryptography, representation theory, benchmarking and industrial use-case of quantum network protocols. We would also like to point out that it is a good tool for students, and all beginners in general, to learn about different protocols and to get them started in the field, thereby broadening the community by providing easy access. The fact that it has already been used in hackathons further supports this observation. Finally, we invite the community to actively submit their new protocols to the zoo given the framework presented in this paper, which will be reviewed by the zoo editors.

\section*{Author Contribution}
EK came up with the idea of the quantum protocol zoo; she provided expert supervision throughout the project. SS was a student intern at LIP6, supervised by EK when she co-founded the quantum protocol zoo with EK. SS, MDo and MDe came up with the idea of the knowledge graph that was implemented by NM and HO from the existing pages on the zoo. MDo and AM extracted the connection between abstract cryptography and the knowledge graph. MDo, SS, HO, NM and MDe wrote the manuscript while EK and AM reviewed it with critical comments.
 
\section*{Acknowledgement}
The \emph{Quantum protocol zoo} project had many collaborators who contributed to different stages of the development of the library. Through all these steps, we had fruitful discussions with quantum groups at TU Delft, LIP6 Sorbonne Université and
School of Informatics at the University of Edinburgh, classical
Internet security at ENS Paris, Security and Privacy Groups
at the University of Edinburgh, VeriQloud and SAP security
group to develop a useful representation of quantum protocols
for each targeted community.
We thank all the zoo contributors who made additions to the zoo apart from the authors up to this date including NM, Rhea Parekh, Bas Dirke, Victoria Lipinska, Gláucia Murta, Jérémy Ribeiro, and Gozde Ustun. We also express our gratitude to our expert reviewers Céline Chevalier (classical cryptography expert), Niraj Kumar (quantum information experiments expert), Kaushik Chakraborty (quantum networks and cryptography expert) and Marc Kaplan (industry expert). Finally, we would like to acknowledge the QOSF mentorship program, as part of which Lucas Arenstein, Isabel Minh Le, Sara Sarfaraz, and Chirag Wadhwa, supervised by SS, MDo and NM, also contributed to the zoo. The zoo was used at several Quantum Internet hackathons as a tool for introduction to quantum communication for classical security experts. We would like to acknowledge all the interns at LIP6 as well as the hackathon teams that contributed towards the code repository on Quantum Protocol Zoo. We give a special mention to Rhea Parekh and Gozde Ustun who created the certification repository and the code repository on the zoo, respectively. Ustun developed this code repository using the pseudo codes from our textual description. Finally, we welcome all researchers to join and add their protocols to the protocol zoo library and keep it alive as a resource for the whole quantum community. 

The authors acknowledge financial support from the ANR International Project VanQute; the European Union’s Horizon 2020 Research and Innovation Programme under Grant Agreement  No. 820445 (QIA); The UK Engineering and Physical Sciences Research Council Grant No. EP/N003829/1; ANR Project SecNISQ;  as well as technical support from VeriQloud for hosting the Zoo and all Paris Center for Quantum Computing members for their generous time in many endless discussions for setting up the zoo. The authors would like to also thank Kaushik Chakraborty and Subhayan Roy Moulik for some critical feedback on the paper.

\bibliographystyle{abbrv}
\bibliography{zoo_ref}
\onecolumn

\appendix
\section{Tableau of Textual Representation}
\begin{table}[!htb]
\centering
\begin{tabular}{|p{0.35\linewidth} | p{0.6\linewidth}|}
    \hline &\\
       Functionality description & A clear articulation of the intended task. \\&\\
       \hline &\\
       Use case  & An assessment of practicality within the industry, as compared to alternative solutions. This aspect is pivotal in recognizing crucial quantum resources and the necessary experimental achievements in that domain. It aims to address the following questions pertinent to the latest implementations and protocol designs\\
       &\tabitem Is it a novel quantum task or a quantum-enhanced implementation of a corresponding classical task?\\
       &\tabitem Are there any classical or post-quantum secure schemes that achieve the same functionality? If yes, how does the quantum solution compare in terms of the efficiency of the latest implementation, scalability, accessibility, security, etc.?\\
       &\tabitem With respect to the previous question, what are the corresponding benchmark values for key length, security parameters, threshold values, etc. from the state-of-the-art implementation?\\&\\
      \hline &\\
      Protocols&It helps in identifying which protocol design can be tested with the available quantum technologies and what its associated state-of-the-art theoretical and experimental advancements are. It presents a list of protocol designs achieving functionality at different stages of the quantum internet~\cite{wehner2018quantum}. Each of these lists comprises a protocol structure explained later.\\&\\
      \hline &\\
      Properties &All the properties that shall be satisfied by any protocol to accomplish the task defined by the functionality, such as,\\
      &\tabitem definitions of security and correctness associated with the quantum functionality set-up;\\
      &\tabitem what are the key requirements for a protocol design to satisfy the definitions and requirements of the target quantum functionality? The section is a quick overview of how the quantum task differs from the existing classical analogues.\\&\\
      \hline &\\
      Further information&Any issues that could not be addressed or find a place in the above sections or any review paper discussing a feature of various types of protocols
  related to the functionality. We compile a list of associated reviews and any connections/use cases of the functionality to other fields in science.\\&\\
  \hline
    \end{tabular}
    \caption{Functionality page format}
    \label{tab:functionality}
\end{table}

\begin{table}[!htb]
    \centering
    \begin{tabular}{|p{0.35\linewidth} | p{0.6\linewidth}|}
    \hline &\\
            Protocol abstract &  It briefly illustrates how the protocol achieves the task.\\&\\
        \hline &\\
        Assumptions & describes the setting in which the protocol is running. Any assumption on the setup or limitation in the actions of parties is listed in this section.\\&\\
        \hline &\\
        Requirements&It specifies the network stage\cite{wehner2018quantum} to which the protocol belongs, relevant network parameters as specified for each stage, and hardware requirements for each party. The associated network stage page lists all the protocols from various functionalities that can be realized at the concerned stage. Details from state-of-the-art experimental implementations such as qubit error rate, number of qubits used, the order of digital threshold values, key length, security parameters and scalability are also detailed in this section. This section is linked to a subgraph of the graphical visualization, the knowledge graph, which links the requirements of the protocols to other protocols lower in the `hierarchy of quantum protocols' (see Fig.~\ref{fig:network-stack} and Sec.~\ref{sec:KGRep}), also identified as the building blocks for the concerned protocol show at the end of this section.\\&\\
        \hline &\\
        Properties&A list of important information extracted from the protocol such as security claims related to the properties given in the root functionality. Results from the theoretical protocol design which claims to achieve the best bounds on various parameters related to correctness and security are specified here.\\&\\
        \hline &\\
        Protocol description&Functional step-wise protocol algorithm helpful to writing the code for the simulation or actual implementation. It can be divided into stages common for all the protocols in the target functionality.\\&\\
        \hline &\\
        Further information&Includes a short description below for a list of all protocols that are variants of the protocol under discussion or are in the same class of functionality and network stage. We specify the differences in the protocol described above and the protocols listed in terms of hardware requirements, properties and security claims. Any new protocol which is easy to interpret after reading the formal description can be included here in the following format.\\&\\
   &\textbf{Theoretical papers}\\&\\
    &\tabitem How is it different from the above protocol?\\
    &\tabitem Which additional Requirements does it need?\\
    &\tabitem Which level of security can it provide?\\
   &\textbf{Experimental papers}\\
    &\tabitem On which platform has the protocol been implemented and which theory design was used?\\
    &\tabitem What benchmarking values were achieved for this demonstration?\\
    \hline
    \end{tabular}
    \caption{Protocol page format}
    \label{tab:protocol}
\end{table}

\newpage
\subsection{Atomic functions}\label{apx:atomic}
\subsubsection{Extracted functions per protocol studied}: Each row of the following table shows the set of protocols that can be used for achieving the same functionality. 

\begin{longtable}{p{0.4\linewidth}|p{0.6\linewidth}}
Protocol & Functionalities used\\
\hline
\endfirsthead
\multicolumn{2}{l}{Continued from previous page} \\
\hline

Protocol & Functionalities used \\

\hline
\endhead
\hline\multicolumn{2}{r}{Continued on next page} \\
\endfoot
\endlastfoot
\hline
GHZ-based Quantum Anonymous Transmission & Classical authenticated channel\\
\url{https://arxiv.org/abs/quant-ph/0409201} & Creation and broadcast of GHZ state\\
 & Classical collision detection protocol\\
 & Single qubit measurement\\
 & Single qubit Hadamard gate\\
 & Local memory\\
 & Teleportation\\
\hline
Verifiable Quantum Anonymous Transmission & Notification (private computation of classical parity, OR, Rand)\\
\url{https://arxiv.org/pdf/1811.04729.pdf} & Single qubit measurements in the equatorial plane\\
 & Local memory\\
 & (Uses GHZ anonymous transmission as a subroutine)\\
\hline
Polynomial Code based Quantum Authentication & Clifford circuits (error correction)\\
\url{https://arxiv.org/pdf/quant-ph/0205128.pdf} & Local memory\\
\hline
Fast Quantum Byzantine Agreement & Creation and broadcast of GHZ state\\
\url{https://dl.acm.org/doi/10.1145/1060590.1060662} & Multipartite Entanglement Verification\\
 & (Uses oblivious common coin)\\
 & (Uses Verifiable Quantum Secret Sharing)\\
\hline
Quantum Bit Commitment & BB84 Encoding of classical data\\
\url{https://arxiv.org/abs/1108.2879} & BB84 Decoding to classical data\\
 & Secure classical channel\\
 & Fast operations to keep the relativistic constraints\\
\hline
Quantum Coin Flipping & \(\pi/9\) single qubit preparation\\
\url{https://arxiv.org/abs/quant-ph/9904078} & Multi qubit POVM\\
\hline
Gottesman and Chuang Quantum Digital Signature & Local memory\\
\url{https://arxiv.org/abs/quant-ph/0105032} & Swap test\\
 & Stabilizer states creation\\
\hline
Prepare and Measure Quantum Digital Signature (QDS) & BB84 Encoding of classical data\\
\url{https://arxiv.org/abs/1403.5551} & BB84 Decoding to classical data\\
\hline
Measurement Device Independent QDS & Classical authenticated channel\\
\url{https://arxiv.org/pdf/1704.07178.pdf} & Measurement Device Independent QKD link\\
 & BB84 Encoding of classical data\\
 & BB84 Decoding to classical data\\
\hline
Multipartite Entanglement Verification & Classical authenticated channel\\
\url{https://www.nature.com/articles/ncomms13251} & Secure classical broadcast\\
 & Common shared randomness\\
 & Local memory\\
 & BB84 Decoding to classical data\\
 & Creation and broadcast of GHZ state\\
\hline
Quantum Fingerprinting & Clifford gates\\
\url{https://arxiv.org/abs/quant-ph/0102001} & Swap test\\
\hline
BB84 & BB84 Encoding of classical data\\
\url{https://core.ac.uk/download/pdf/82447194.pdf} & BB84 Decoding to classical data\\
 & Authenticated classical channel\\
 & Privacy amplification\\
 & Information reconciliation\\
\hline
Device Independent QKD & EPR distribution\\
\url{https://arxiv.org/abs/1811.07983} & Information reconciliation\\
 & Privacy amplification\\
\hline
Quantum Leader Election & (Uses Weak coin flipping)\\
\url{https://arxiv.org/abs/0910.4952} & \\
\hline
Quantum Cheque & Creation and broadcast of GHZ state\\
\url{https://link.springer.com/article/10.1007/s11128-016-1273-4} & Local memory\\
 & Quantum 1-way function\\
 & SWAP test\\
 & (Uses QKD)\\
\hline
Quantum Coin & Clifford gates\\
\url{http://users.math.cas.cz/\~gavinsky/papers/QuMoClaV.pdf} & Local memory\\
\hline
Quantum Token & BB84 Encoding of classical data\\
 & BB84 Decoding to classical data\\
 & Local memory\\
\hline
Wiesner Quantum Money & BB84 Encoding of classical data\\
\url{http://users.cms.caltech.edu/\~vidick/teaching/120\_qcrypto/wiesner.pdf} & BB84 Decoding to classical data\\
 & Local memory\\
\hline
Quantum Oblivious transfer & BB84 Encoding of classical data\\
\url{https://link.springer.com/chapter/10.1007/3-540-46766-1\_29} & BB84 Decoding to classical data\\
\hline
Classical FHE for Quantum Circuits & Full QC\\
\url{https://arxiv.org/abs/1708.02130} & \\
\hline
Measurement-Only Universal Blind Quantum Computation & Graph state generation\\
\url{https://journals.aps.org/pra/abstract/10.1103/PhysRevA.87.050301} & Equatorial plane measurements\\
\hline
Prepare-and-Send Quantum Fully Homomorphic Encryption & Full QC (server)\\
\url{https://arxiv.org/abs/1603.09717} & Quantum OTP (client)\\
\hline
Prepare-and-Send Universal Blind Quantum Computation & Graph state generation\\
\url{https://arxiv.org/abs/0807.4154} & Equatorial plane measurements\\
\hline
Pseudo-Secret Random Qubit Generator & Full QC on server's side\\
\url{https://arxiv.org/abs/1802.08759} & Quantum-safe one-way functions\\
\hline
Prepare-and-Send Verifiable Universal Blind Quantum Computation & Graph state generation\\
\url{https://arxiv.org/abs/1203.5217} & Equatorial plane measurement\\
 & Quantum One Time Pad\\
 & Local memory\\
\hline
Measurement-Only Verifiable Universal Blind Quantum Computation & Graph state generation\\
\url{https://arxiv.org/abs/1208.1495} & Equatorial plane measurement\\
 & Local memory\\
\hline
Prepare-and-Send Verifiable Quantum Fully Homomorphic Encryption & Full QC (server)\\
\url{https://arxiv.org/abs/1708.09156} & Clifford QC (client)\\
\hline
Secure Multiparty Delegated Quantum Computation & Graph state generation\\
\url{https://arxiv.org/abs/1606.09200} & Verifiable secret sharing\\
\hline
State Teleportation & EPR state source and broadcasting\\
\url{https://doi.org/10.1103/PhysRevLett.70.1895} & BB84 Decoding to classical data\\
\hline
Weak String Erasure & BB84 Encoding of classical data\\
\url{https://eprint.iacr.org/2005/291.pdf} & BB84 Decoding to classical data\\
\end{longtable}

\subsubsection{Alternate graphical representation for the relations between atomic functions and protocols of the zoo}

The graph presented below contains the various protocols and atomic functions extracted from the quantum protocol zoo. The larger nodes are atomic functions, while smaller ones correspond to application-level protocols. The color of the nodes corresponds to the network stage of the task: 
\begin{itemize}
\item dark green: classical
\item green: prepare and measure
\item yellow: trusted repeater
\item orange: entanglement distribution
\item red: quantum memory
\item purple: quantum computing.
\end{itemize}

This graph clearly shows the importance of BB84 encoding as well as Sending qubit and Local memory functionalities as measured by the degree-centrality. Less apparent without this graph are the role played by GHZ and graph state preparations. Their importance stems from their usefulness in most delegated quantum computing schemes as well as multipartite protocols. This would call for specific attention in implementing these functionalities, maybe even considering dedicated network architectures adapted to the production of these resource states.

Here, note that we did not represent possible implementations of atomic tasks. Instead, atomic functions are kept \emph{atomic} because a single atomic function could be implemented in various ways each using different functionalities (see eg. creation and distribution of GHZ states involving broadcasting EPR pairs vs. local creation and qubit sending). Opting for one implementation against another is not the purpose of this report as flexibility should be in the hands of experimentalists in order for them to optimize the quality of the produced functionality given its input/output specification.

\begin{center}
\includegraphics[width=.9\linewidth]{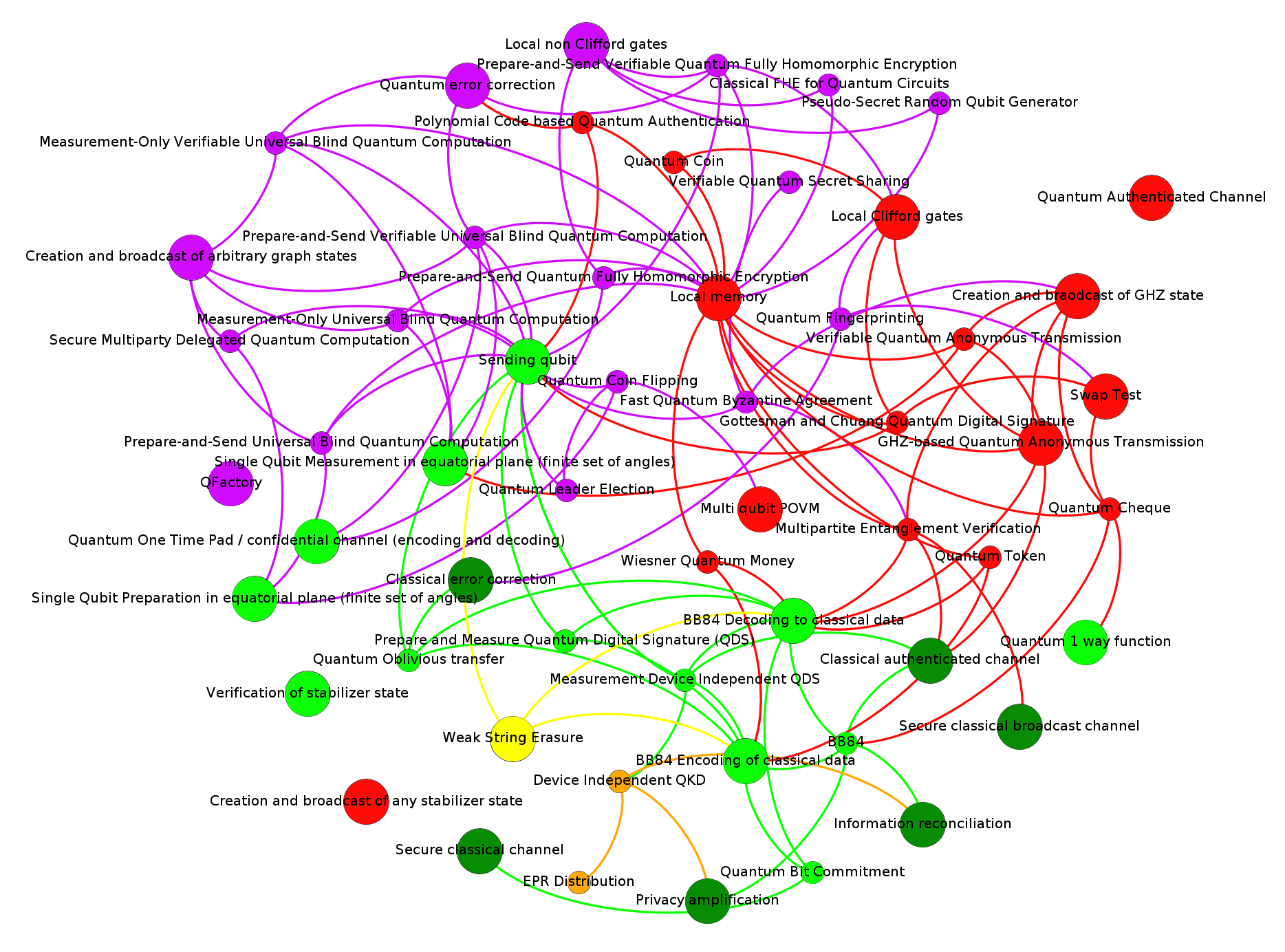}
\end{center}

\vfill

\end{document}